\documentclass[10pt]{iopart}

\usepackage{lipsum,capt-of,graphicx}
\usepackage[squaren]{SIunits}
\usepackage{verbatim}
\usepackage{mathrsfs}
\usepackage{iopams}
\usepackage{comment}
\usepackage{braket}
\usepackage{bbold}
\usepackage{subfigure}
\usepackage{dcolumn}
\usepackage{hyperref}
\usepackage{enumerate}
\usepackage[11pt]{moresize}
\usepackage[normalem]{ulem}
\usepackage{xcolor}
\usepackage{color}
\usepackage{times}
\usepackage{amssymb}
\usepackage{booktabs}
\newcommand{\ra}[1]{\renewcommand{\arraystretch}{#1}}

\renewcommand{\thesubfigure}{\thefigure.\arabic{subfigure}} \makeatletter
\renewcommand{\p@subfigure}{}
\renewcommand{\@thesubfigure}{\thesubfigure.\,\,\hskip\subfiglabelskip} \makeatother

\newcommand{\refssyrho}{(6)\,}

\begin{document}

\title{Spatial correlations in driven-dissipative photonic lattices}

\author{Matteo Biondi}
\address{Institute for Theoretical Physics, ETH Zurich, 8093 Z\"urich, Switzerland}
\author{Saskia Lienhard} 
\address{Institute for Theoretical Physics, ETH Zurich, 8093 Z\"urich, Switzerland}
\author{Gianni Blatter}
\address{Institute for Theoretical Physics, ETH Zurich, 8093 Z\"urich, Switzerland}
\author{Hakan E. T\"ureci}
\address{Department of Electrical Engineering, Princeton University, 08544 Princeton, 
New Jersey, USA}
\author{Sebastian Schmidt}
\address{Institute for Theoretical Physics, ETH Zurich, 8093 Z\"urich, Switzerland}

\begin{abstract}
We study the nonequilibrium steady-state of interacting photons in cavity arrays as described by the driven-dissipative
Bose-Hubbard and spin-$1/2$ XY model. For this purpose, we develop a self-consistent expansion in the inverse coordination number of the array 
($\sim 1/z$) to solve the Lindblad master equation of these systems beyond the mean-field approximation. Our formalism is compared
and benchmarked with exact numerical methods for small systems based on an exact diagonalization of the Liouvillian and a recently developed corner-space renormalization technique. 
We then apply this method to obtain insights beyond mean-field in two particular settings: (i) We show that the gas--liquid transition in the driven-dissipative Bose-Hubbard model is characterized by large density fluctuations and bunched photon statistics. (ii) We 
study the antibunching--bunching transition of the nearest-neighbor correlator in the driven-dissipative 
spin-$1/2$ XY model and provide a simple explanation of this phenomenon.
\end{abstract}

\maketitle

\section{Introduction\label{sec1}}

In recent years interacting photonic lattices have emerged as a versatile platform for the study of many-body phenomena out
of equilibrium \cite{houck2012,schmidt2013*2,carusotto2013, Noh2016, Hartmann2016}. First prototype quantum simulators have been realized experimentally based on 
cavity and circuit QED technologies \cite{hafezi2013,Raftery2014,Eichler2015,Baboux2015,Anderson2016,Fitzpatrick2017,Fink2017}. The 
increasing experimental interest in assembling cavities to form lattices is also a strong motivation to develop novel theoretical tools.
The key object governing the dynamics of such driven-dissipative systems is typically the Liouvillian superoperator \cite{scully1997}, which describes the dynamical evolution of the system density matrix $\rho$ 
through a master equation. Solving the master equation exactly is a formidable numerical task \cite{breuer2002}. While exact diagonalization and quantum-trajectory algorithms \cite{Dalibard1992,carmichael1992,Plenio1998,Daley2014} 
allow to successfully address this problem for small system sizes, large scale numerical methods based on matrix-product-states (MPS) \cite{zwolak2004,Schollwoek2011,Cui2015,dorda2015,Mascarenhas2015,biondi2015}
are typically limited to one dimension (1D). Recently developed methods such as the corner-space renormalization technique \cite{Finazzi2015} may provide a promising alternative also in two dimensions (2D).
On the other hand, decoupling mean-field theory, which is correct in infinite lattice dimensions, is a simple yet valuable tool to gain a first insight into the qualitative physics at work. It has been successfully applied to various
lattice models such as the Bose-Hubbard and Jaynes-Cummings-Hubbard model \cite{nissen2012,jin2013,boite2013,Biondi2016,FossFeig2017,Biella2017} as well as related spin models \cite{Lee2012,Ates2012, Wilson2016}. 
Recent efforts to improve on the mean-field approximation include perturbative \cite{delValle2013,Li2014}, 
projective \cite{DegenfeldSchonburg2014}, cluster \cite{jin2016}, variational \cite{Weimer2015} and equations-of-motion approaches \cite{Casteels2016}. 

Here, we develop a systematic expansion around the decoupling mean-field solution of the Lindblad master equation 
in powers of the inverse dimensionality parameter $1/z$ (with $z$ being the number of nearest neighbors in a lattice).
Such an expansion accounts for quantum fluctuations in a systematic way and provides access to a whole new class of observables, i.e., spatial correlation functions. For systems in (quasi-) equilibrium, 
which are fully described by the Hamiltonian alone, the $1/z$ 
expansion has a diagrammatic interpretation in terms of linked-clusters and was used to calculate the ground-state and elementary excitations of 
Fermi-Hubbard \cite{Metzner1991}, Bose-Hubbard \cite{Ohliger2013} and Jaynes-Cummings-Hubbard \cite{schmidt2009} models. In the nonequilibrium context, this technique was employed 
in Refs.~\cite{Navez2010,Queisser2012} to calculate quenched dynamics of atoms in optical lattices
and in Ref.~\cite{Weimer2015*2} to characterize the transition from low to high density phases in a driven, dissipative Rydberg system.

In this work, we expand on previous efforts by developing a method to solve for the density matrix in a self-consistent way.
We calculate the nonequilibrium steady-state of the driven-dissipative Bose-Hubbard model up to second order in $1/z$
and show that the self-consistency condition substantially improves the results by comparing to exact diagonalization in 1D and the 
corner-space method in 2D. We then apply our method to two specific problems: 
(i) we calculate the compressibility of the  driven-dissipative Bose-Hubbard model
and show that the photonic gas--liquid transition is characterized by largely enhanced density fluctuations with bunched photon statistics;
(ii) we study the antibunching--bunching transition of the driven-dissipative spin-1/2 XY model
in one and two dimensions and provide a simple explanation based on a dimer model.

The remainder of the paper is structured as follows. In Section~\ref{sec2}, we introduce two models 
for interacting photons in cavity arrays, the driven dissipative Bose-Hubbard and the spin-$1/2$ XY model. 
In Section~\ref{sec3}, we discuss the self-consistent $1/z$ expansion and benchmark our method 
by comparing with numerical results based on exact diagonalization and the corner-space renormalization technique. 
In Section \ref{sec4}, we address the effects of site-site correlations in the gas--liquid transition of the driven-dissipative Bose-Hubbard model. 
In Sections~\ref{sec5}, we study the driven-dissipative spin-1/2 XY model
to discuss the antibunching--bunching transition in one and two dimensions. 
In Section \ref{sec7} we summarize the results of the paper and provide an outlook for future work.

\section{Model\label{sec2}}
We investigate the steady-state of the coherently pumped and
dissipative Bose-Hubbard model (BHM) describing photons hopping on a lattice
of nonlinear cavities with local coherent pump and decay. The lattice
Hamiltonian reads
\begin{eqnarray}
\label{h_BHM}
   H & = \sum_i h_i + \frac{1}{z} 
   \sum_{ \langle ij \rangle } J_{ij} a^\dagger_i a_j, \\
      h_i & = - \Delta\,n_i + Un_i(n_i - 1)/2 
   + f(a_i + a_i^\dagger).
\end{eqnarray}
Here, each site $i$ is coherently pumped with strength $f$ as described by the
last term in $h_i$, which is expressed in terms of
the bosonic operator $a_i$ and the associated density operator $n_i =
a^\dagger_i a_i$. In a frame rotating with the drive frequency $\omega_d$ the
cavity frequency is renormalized to $\Delta = \omega_d - \omega_c$, while $U$
is the local Kerr nonlinearity. The second
term in $H$ describes the hopping to $z$ nearest-neighbor cavities with
amplitude $J_{ij} = -J$; the additional factor $1/z$ in~\eref{h_BHM}
ensures that the bandwidth of the photon dispersion is $2J$, independent of $z$,
and guarantees a regular limit $z\to \infty$. The dissipative dynamics for the
density matrix $\rho$ is accounted for via Lindblad's master equation,
\begin{eqnarray}
\dot{\rho} = -i[H,\rho] + \frac{\kappa}{2} \sum_i D[a_i] \rho,
\label{lindblad_full}
\end{eqnarray}
where $D[a] \rho = 2 a\rho a^\dagger - a^\dagger a\rho - \rho a^\dagger a$ and
$\kappa$ is the photon decay rate. This model can be realized in quantum
engineered settings using state-of-the-art superconductor-
\cite{houck2012,schmidt2013*2} as well as semiconductor technologies
\cite{carusotto2013}. In the limit of large on-site nonlinearity ($U\rightarrow \infty$), the
double occupation of lattice sites is suppressed and the local Hilbert space
cutoff $n_p$ (i.e., the maximal number of photons per site) can be restricted
to unity ($n_p = 1$). In this regime, photon operators are mapped to spin Pauli
operators $a_i \rightarrow \sigma_i^-$ with corresponding ground $\ket{g_i} = \ket{0_i}$
and excited state $\ket{e_i} = \ket{1_i}$, where $\ket{0_i} (\ket{1_i})$ denote photon Fock states with zero (one) photons 
at site $i$. Consequently, the BHM becomes equivalent to the
spin-$1/2$ XY model (XYM) with the Hamiltonian
\begin{eqnarray}
\label{h_XYM}
   H & = \sum_i h_i + \frac{1}{z} 
   \sum_{\langle ij \rangle} J_{ij} \sigma^+_i \sigma^-_j,\\
   h_i & = -\Delta\, n_i 
   + f(\sigma_i^+ + \sigma_i^-).
\end{eqnarray}
Here, dissipation is taken into account as in \eref{lindblad_full} with the
collapse operator replacement $a_i \rightarrow \sigma_i^-$. 

\section{Expansion in $1/z$ and benchmarking \label{sec3}}

In the following, we describe a strong coupling expansion in powers of the inverse
coordination number $z$, which was originally developed to calculate the ground-state properties and elementary excitations 
of various Hubbard-type lattice models under equilibrium conditions \cite{Metzner1991,Ohliger2013,schmidt2009}.
Recently, such a $1/z$ expansion was also carried out in the nonequilibrium context to study
quenched dynamics of atoms in optical lattices \cite{Navez2010,Queisser2012} as well as dissipative Rydberg gases \cite{Weimer2015*2}. 
Here, we will expand on these early efforts in the nonequilibrium context and develop a self-consistent scheme, which is correct to second order in $1/z$.
We will show that self-consistency considerably improves the mean-field approximation. It allows to systematically account for quantum fluctuations yielding quantitatively correct results in a large parameter range
even for small lattice sizes. While we focus here on the BHM and XYM, the technique is rather generic and applicable to 
a wide range of driven, dissipative lattice models with limited range hopping.

We start by defining the reduced density matrices of
one lattice site $\rho_i = \tr_{\neq i} [\rho]$, two lattice sites $\rho_{ij}
= \tr_{\neq ij} [\rho]$, three lattice sites $\rho_{ijk} = \tr_{\neq ijk}
[\rho]$, etc. The trace $\tr_{\neq i,\dots, n}$ sums over all photon states of
all cavities except those indexed with the subscript.  The few-site density
matrices $\rho_{i, \dots, n}$ are represented in photon number space and their
matrix elements read, e.g., $ \rho_{n_im_i} = \bra{n_i}\rho_i \ket{m_i}$, $
\rho_{n_im_i p_jq_j} = \bra{n_i p_j}\rho_{ij} \ket{m_iq_j}$, where
$\ket{n_i},\ket{p_j}$ etc. denote photon number states at
site $i$, $j$ etc.  These density operators can be decomposed into connected
and factorizable terms, i.e., $\rho_{ij} =  \rho_{ij}^c +
\rho_{i}\rho_{j}$, $\rho_{ijk} =  \rho_{ijk}^c + \rho_{ij}^c\rho_{k} +
\rho_{ik}^c\rho_{j} + \rho_{jk}^c\rho_{i} + \rho_{i}\rho_{j} \rho_{k}$, etc.
A systematic expansion in powers of $1/z$ can then be organized based on the hierarchy of correlations $\rho_{i_1,i_2,\dots,i_s}^c = \mathcal{O}(1/z^{s-1})$, where
$s$ is the number of lattice sites in the connected density matrix $\rho_{i_1,i_2,\dots,i_s}^c$. 
In particular, $\rho_i$ is of order unity, i.e., $\rho_i = \mathcal{O}(1/ z^0)$,
$\rho_{ij}^c$ is of order $1/z$, and $\rho_{ijk}^c$ is of order $1/z^2$.
Such a scaling of correlations is known from the Bogoliubov-Born-Green-Kirkwood-Yvon (BBGKY) hierarchy of statistical mechanics \cite{Cercignani1997}, with the difference that it here applies to lattice sites instead of particles.  
Starting from~\eref{lindblad_full}, we obtain the equation of motion for the reduced density matrices up to order $1/z^2$ \cite{Navez2010,Queisser2012}, i.e.,

\numparts
\begin{eqnarray}
\label{sys_loc}
i\dot{\rho}_i = & \mathcal{L}_i \rho_i + \frac{1}{z}\sum_{j\neq i} 
\tr_{j}[\mathcal{L}^{\rm \scriptscriptstyle S}_{ij} (\rho_{i} \rho_{j} +  \underline{\rho^c_{ij}})],\\
i\dot{\rho}_{ij}^c = & \mathcal{L}_i \rho_{ij}^c 
+ \frac{1}{z}\mathcal{L}_{ij} (\rho_{i}\rho_{j} + 
\underline{\rho^c_{ij}}) - \frac{\rho_i}{z}
\tr_i[\mathcal{L}^{\rm \scriptscriptstyle S}_{ij} (\rho_{i}\rho_j + \underline{\rho^c_{ij}})] \nonumber\\
& + \frac{1}{z}\sum_{k\neq ij} \tr_k[\mathcal{L}^{\rm \scriptscriptstyle S}_{ik}(\rho_{ij}^c\rho_{k} + \rho_{jk}^c \rho_{i} + \underline{\rho_{ijk}^c})] 
+ (i \leftrightarrow j), \label{sys_conn}\\
\label{sys_conn_3sites}
i\dot{\rho}_{ijk}^c = & \mathcal{L}_i \rho_{ijk}^c +  
\frac{1}{z}\mathcal{L}^{\rm \scriptscriptstyle S}_{ij} (\rho_{ik}^c\rho_j + \rho_{jk}^c\rho_i) 
\nonumber\\
& - \frac{\rho_i}{z}\tr_i[\mathcal{L}^{\rm \scriptscriptstyle S}_{ij}(\rho_{ik}^c\rho_j + \rho_{jk}^c\rho_i + \underline{\rho_{ijk}^c}) 
+\mathcal{L}^{\rm \scriptscriptstyle S}_{ik}
(\rho_{ij}^c\rho_k + \rho_{jk}^c\rho_i + \underline{\rho_{ijk}^c})] \nonumber \\
& - \frac{\rho_{ik}^c}{z}\tr_i[\mathcal{L}^{\rm \scriptscriptstyle S}_{ij}(\rho_{i}\rho_j  
+ \underline{\rho_{ij}^c})] - \frac{\rho_{ij}^c}{z}
\tr_i[\mathcal{L}^{\rm \scriptscriptstyle S}_{ik}(\rho_{i}\rho_k  + 
\underline{\rho_{ik}^c})] \nonumber\\
& + \frac{1}{z}\sum_{k'\neq ijk}\tr_{k'}
[\mathcal{L}^{\rm \scriptscriptstyle S}_{ik'}(\rho_{ijk}^c\rho_{k'}  
+ \rho_{jkk'}^c\rho_{i}  + \rho_{ij}^c\rho^c_{kk'} + \rho_{ik}^c\rho^c_{jk'} \nonumber \\ 
& + \underline{\rho_{ijkk'}^c})] + ( i \to j,j \to k, k \to i) +  ( i \to k,j \to i,k \to j).
\end{eqnarray}
\endnumparts
Above, we introduced the notation $\mathcal{L}^{\rm \scriptscriptstyle S}_{ij} = \mathcal{L}_{ij}
+ \mathcal{L}_{ji}$, $\mathcal{L}_{ij} \rho = J_{ij}[a_i^\dagger a_j, \rho]$
and $\mathcal{L}_i\rho = [h^{\scriptscriptstyle \rm BH}_i,\rho] + i(\kappa/2)D[a_i]\rho$ as in Refs.~\cite{Navez2010,Queisser2012}. 
In the mean-field limit of infinite coordination number ($z\rightarrow\infty$) all connected density matrices are zero and 
one only needs to solve~\eref{sys_loc}, which is nonlinear and can have multiple solutions. 
However, in order to account for spatial correlations, one needs to evaluate the density matrix to higher order in $1/z$ and also
solve the equations of motion for the connected density matrices. In a first step, we make use of the scaling hierarchy $\rho_{i_1,i_2,\dots,i_s}^c = \mathcal{O}(1/z^{s-1})$
and keep on the r.h.s of each equation only terms up to order $1/z^{s-1}$, where $s$ is the number of lattice sites
in the connected density matrix on the l.h.s.~of each equation (i.e., we neglect the underlined terms). 
The resulting system of equations is then closed and 
can be solved numerically. Note, that in this case the equations for the connected density matrices are linear and depend
on the solution of the nonlinear mean-field equation only parametrically. 

\begin{figure}[t]
\centering
\includegraphics[width=0.6\textwidth]{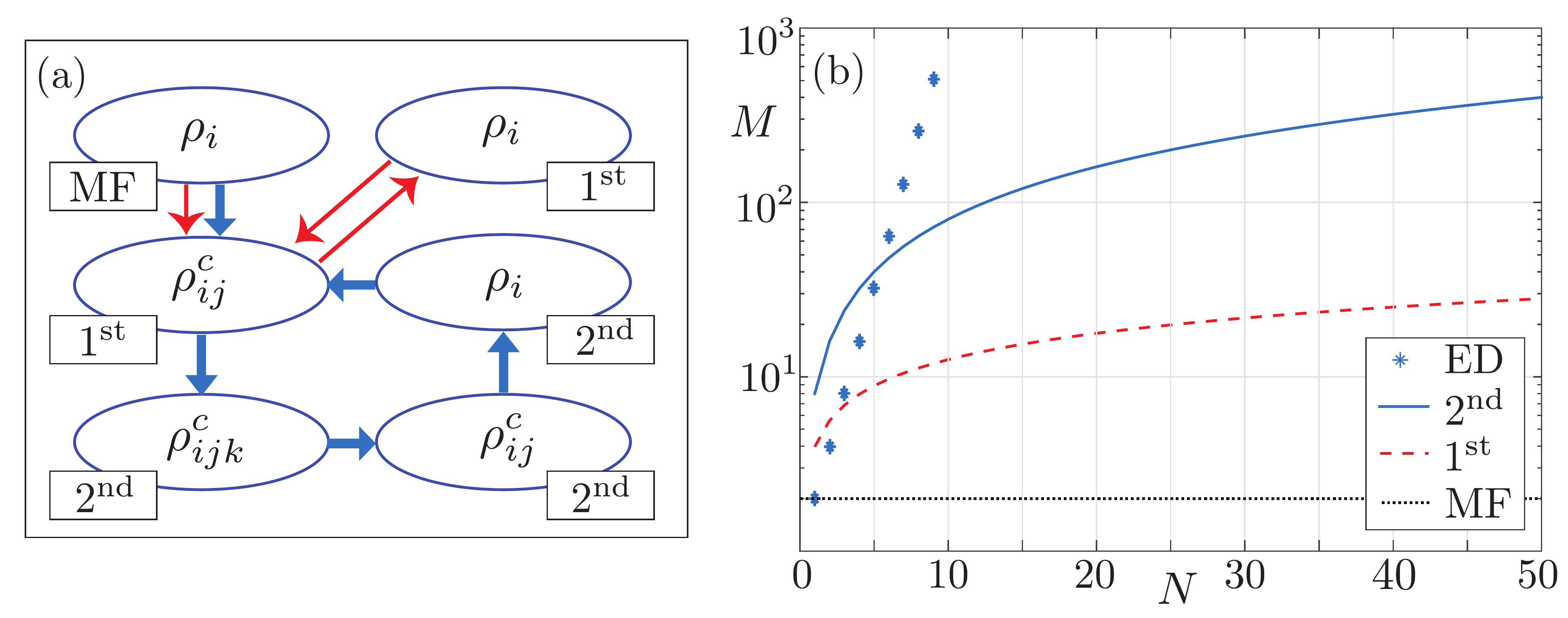}
\caption{(a) Illustration of the self-consistent scheme 
explained in detail in the text in order to solve
the equation of motion system \refssyrho{}. The solid blue (thin red) arrows indicate the
procedure to obtain a lattice density matrix correct to second, $2^{\rm nd}$
(first, $1^{\rm st}$) order in $1/z$.  (b) Scaling of the dimension of the
Liouvillian operator $M^2\!\times M^2$ with the number of
lattice sites $N$, for the $1/z$ expansion (curves) and exact diagonalization
(ED, symbols) for a system with maximally one particle per site $n_p =1$. The
scaling for larger $n_p$ is similar. \label{fig_scheme}}
\end{figure}

In the following, we substantially improve this first approximation by keeping explicitly all underlined terms to second order in $1/z$ in the system of equations above, i.e., by
neglecting only the third order term in \eref{sys_conn_3sites} ($\sim \rho_{ijkk'}^c$). We then solve the coupled system of equations in a self-consistent way taking the following steps:
(i) we solve \eref{sys_loc} for $\rho_i$; (ii) the result
is inserted into \eref{sys_conn} to obtain $\rho_{ij}^c $; (iii) $\rho_i$ and
$\rho_{ij}^c$ are used to solve \eref{sys_conn_3sites} for $\rho_{ijk}^c$. Note, that
(i)-(iii) correspond to the first step, which was explained in the previous paragraph. 
In order to implement self-consistency we now explain the second step, i.e., 
(iv) insert $\rho_{ijk}^c$ back in \eref{sys_conn} and obtain an updated
$\rho_{ij}^c$; (v) plug $\rho_{ij}^c$ in \eref{sys_loc} and get a new
$\rho_i$.  In (iv-v) all the underlined terms are kept. Starting from the
updated $\rho_i$, the procedure (ii-v) is iterated till convergence is
reached. This yields a solution of the hierarchy equations \refssyrho{}
correct to second ($2^{\rm nd}$) order $1/z^2$, i.e., with an error on the
density matrix of order $\mathcal{O}(z^{-3})$. Without the steps (iii-iv) the solution of the hierarchy equations is correct
to first ($1^{\rm st}$) order $1/z$, i.e., with an error on the density matrix
of order $\mathcal{O}(z^{-2})$. Step (i) alone is correct to zeroth order and equivalent to a
Gutzwiller mean-field (MF) decoupling of the hopping term in the Hamiltonian
\eref{h_BHM}. The sequence of steps performed in this self-consistent scheme
is illustrated in figure~\ref{fig_scheme}(a).

In figure~\ref{fig_scheme}(b) we show how the numerical complexity of the method
scales with the number of lattice sites $N$ and compare to an exact numerical solution of
the master equation. The size of the full Hamiltonian is given by $M = (n_p
+ 1)^N$ and thus increases exponentially with the number of lattice sites (see blue dots in figure~\ref{fig_scheme}(b)).
An exact solution of the master equation is then obtained by writing the density matrix as a vector $\brho$
of density matrix elements with length $M^2$ such that~\eref{lindblad_full} can be rewritten as 
$\boldsymbol{\dot{\rho}} = \mathcal{L}\boldsymbol{\rho}$, where $\mathcal{L}$ is the corresponding
Liouvillian operator with dimension $M^2\times M^2$. A diagonalization of the 
(non-hermitian) Liouvillian then yields in general complex eigenvalues and eigenfunctions which fully determine the exact solution.

Let us now estimate the computational complexity of the $1/z$ expansion assuming a translationally invariant density matrix 
with periodic boundary conditions. In this case, the solution of the nonlinear equation \eref{sys_loc} remains site-independent, even when the underlined term is included, i.e., its complexity does not substantially depend on the number of lattice sites. The remaining linear system of equations, which has to be solved iteratively, takes the form $ \boldsymbol{\dot{\rho}}_c = \mathcal{L}\boldsymbol{\rho}_c + \boldsymbol{b}$, where $\mathcal{L}$ is again the
corresponding Liouvillian-type operator, $\boldsymbol{b}$ is a source term, and the
vector $\boldsymbol{\rho}_c$ contains the matrix elements of the connected
density matrices $\rho^c_{ij}$ and $\rho^c_{ijk}$.
Here, the length of the vector $\boldsymbol{\rho}_c$ is $M^2$ with $M = N^{\lambda/2} \times (n_p + 1)^{\lambda +1}$,
where $\lambda=1,2$ corresponds to the order of the expansion. 
Consequently, the computational effort scales only polynomially with the number of lattice sites.
In figure~\ref{fig_scheme}(b), 
we compare the dimensions of the Liouvillian operators for
$n_p=1$. For example, the calculation of the XYM on a square lattice
with $7\times 7$ sites would involve a very large Liouvillian operator with
$M \approx 10^{15}$, which would be far beyond sparse ED methods and even stochastic
techniques based on quantum trajectories \cite{Dalibard1992} where $M\approx10^6$ forms an upper limit. 
On the other hand, the density matrix of such a large system can be easily
computed using the coordination number expansion to second order even on a standard laptop
computer (e.g., see results in table~\ref{tables}).

\begin{table}
\tabcolsep=0.1cm
\ra{1.5}
\caption{\label{tables} Density $n$, on-site and nearest-neighbor correlators $g^{\scriptscriptstyle (2)}_{0j}$
with $j=0,1$ for the BHM. Shown are results obtained from the $1/z$ expansion
applied to a 1D array with 6 sites and cutoff $n_p=2$ (a) and to a 2D square lattice
(b) with $4\times4$ sites ($U/\kappa=20$, $n_p=3$), $3\times3$ sites
($U/\kappa=10$, $n_p=5$), and $7\times7$ sites ($U/\kappa=1$, $n_p=4$), where
$n_p$ is the local photon cutoff. The $1/z$ results are compared with ED in
(a) and with data from the corner-space method (CM) \cite{Finazzi2015} in (b).
We have used the parameters $\Delta/\kappa=5$, $f/\kappa=2$, and $J/ \kappa=1$.
The rows marked with an asterisk $\ast$ show results for a large hopping
$J/\kappa=3$ where the agreement is less favorable.}
\begin{indented}
\item[]\begin{tabular}{@{}lrrrrlrrrrlrrr@{}} 
\br
(a) 1D & \multicolumn{4}{c}{$n$} & & \multicolumn{4}{c}
{$g^{\scriptscriptstyle (2)}_{00}$} && 
\multicolumn{3}{c}{$g^{\scriptscriptstyle (2)}_{01}$}
\\ 
\cline{2-5} \cline{7-10} \cline{12-14}
$U/\kappa$ & MF & $1^{\rm st}$ & $2^{\rm nd}$ & ED &  & MF & $1^{\rm st}$ & $2^{\rm nd}$ & ED  & & 
$1^{\rm st}$ & $2^{\rm nd}$ & ED\\
$1$  & 0.113 & 0.113 & 0.113 & 0.113 & & 1.015 & 1.006 & 1.008 & 1.008 & & 1.018 & 1.027 & 1.026\\
$10$  & 0.850 & 0.820 & 0.823 & 0.823 &  &0.651 & 0.672 & 0.669 & 0.669 & & 0.971 & 0.973 & 0.972\\
$20$  & 0.123&  0.128 & 0.130 & 0.130 && 0.815 & 0.850& 0.869 & 0.869 & & 1.338 & 1.425 & 1.420\\
$20^*$  & 0.076 & 0.104 & 0.148 & 0.137 & & 0.870 & 1.111 & 1.226 & 1.257 & & 1.986 & 2.210 & 2.241\\
\br
\end{tabular}
\end{indented}

\begin{indented}
\item[]\begin{tabular}{@{}lrrrrlrrrrlrrr@{}} 
\br
(b) 2D & \multicolumn{4}{c}{$n$} & & \multicolumn{4}{c}
{$g^{\scriptscriptstyle (2)}_{00}$} && 
\multicolumn{3}{c}{$g^{\scriptscriptstyle (2)}_{01}$}
\\ 
\cline{2-5} \cline{7-10} \cline{12-14}
$U/\kappa$ & MF & $1^{\rm st}$ & $2^{\rm nd}$ & CM &  & MF & $1^{\rm st}$ & $2^{\rm nd}$ & CM  & & 
$1^{\rm st}$ & $2^{\rm nd}$ & CM\\
$1$  & 0.116 & 0.116 & 0.116 & 0.116 & &1.265 & 1.259 & 1.259 & 1.259 & & 0.989 & 0.990 & 0.990\\
$10$  & 0.959 & 0.930 & 0.932 & 0.928 & &0.609 & 0.624 & 0.623 & 0.617 & & 1.007 & 1.008 & 1.007\\
$20$  & 0.125&  0.128 & 0.128 & 0.128 & & 0.839 & 0.853 & 0.860 & 0.860 && 1.173 & 1.172 & 1.172\\
$20^*$  & 0.077 & 0.089 & 0.099 & 0.099 & &0.888 & 1.052 & 1.170 & 1.179 & & 1.521 & 1.715 & 1.63\\
\br
\end{tabular}
\item[] $^{*}$ results obtained with $J/\kappa=3$.
\end{indented}
\end{table}

In table~\ref{tables}, we compare the $1/z$ expansion with (i) the exact diagonalization method (ED) for a 1D chain and (ii) with numerical data available for a 2D square lattice from the so-called corner-space renormalization method (CM) developed in Ref.~\cite{Finazzi2015}. The CM is a numerical algorithm which uses the exact solution of the 
master equation for a small lattice and extrapolates it to larger system sizes. At each extrapolation step of the algorithm, two small lattices are merged to form a
larger one, while truncating the basis of the joint Hilbert space to a small number of most probable states (i.e., the corner-space). For better comparison, we chose the same parameters as in Ref.~\cite{Finazzi2015} for both dimensions. Shown are results for the photon density
\begin{eqnarray}
\label{eq:n}
n = \langle a_i^\dagger a_i\rangle=\langle a^\dagger a\rangle 
\end{eqnarray}
and the second-order coherence (density-density correlator) 
\begin{eqnarray}
\label{eq:g2}
g^{\scriptscriptstyle (2)}_{ij}(t=0) = \frac{\langle a_i^\dagger a_j^\dagger a_ia_j\rangle}{n^2}
\end{eqnarray}
describing instantaneous (zero time delay) correlations between sites $i$ and $j$. The latter is measurable in a 
Hanbury Brown-Twiss setup \cite{Brown1956,Oehri2015}. The average in \eref{eq:n} and \eref{eq:g2} is taken with respect to the
nonequilibrium steady-state (NESS) of equation~\refssyrho with $\dot\rho=0$.

\begin{figure}[t]
\centering
\includegraphics[width=0.65\textwidth]{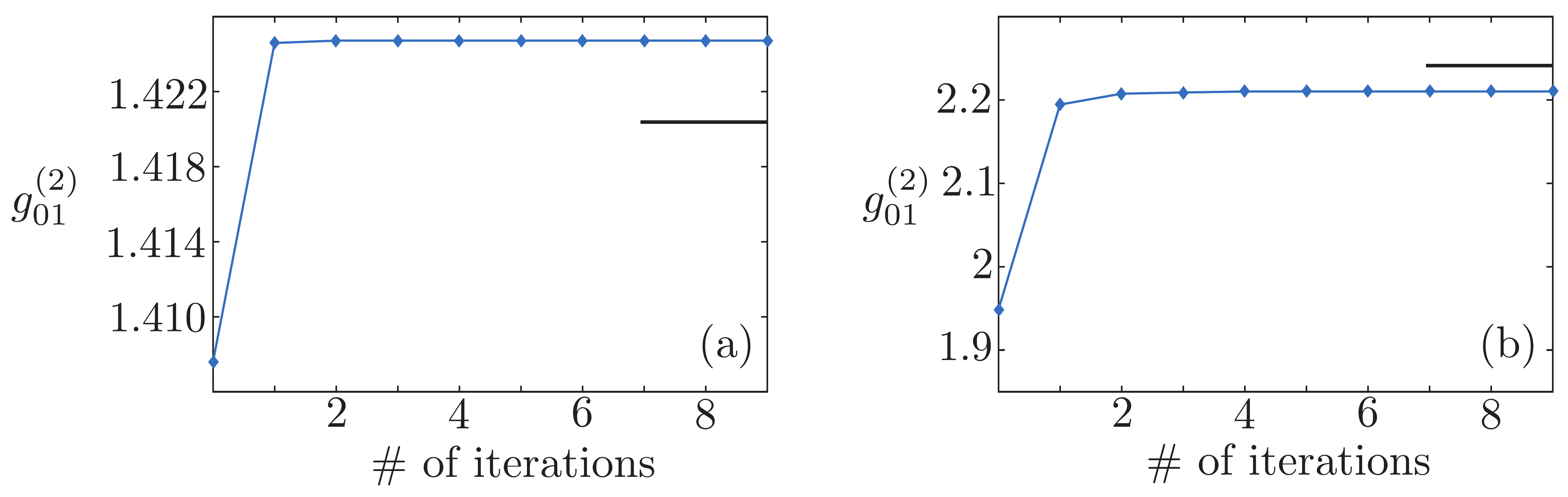}
\caption{Convergence of the nearest-neighbor
correlator $g^{\scriptscriptstyle (2)}_{01}$ as a function of the number
of iterations for the self-consistent scheme to second order in the $1/z$ expansion,
for the third (a) and fourth (b) row in table I(a). The horizontal solid lines indicate the exact value.
\label{convplots}}
\end{figure}

For the parameters considered in table~\ref{tables}, our self-consistent $1/z$ expansion improves the mean-field result (MF) substantially and agrees well with
the exact numerical findings in both dimensions.
In 1D, we find quantitative agreement with the exact result  up to the second and the third decimal for weak to moderate hopping rates ($J\sim\kappa$).
Small discrepancies start to show up for larger hopping rates ($J\sim3\kappa$, see rows marked with an asterisk $\ast$ in table~\ref{tables}) and strong site to site correlations ($g^{\scriptscriptstyle (2)}_{01}\sim 2$).
Such a behaviour is expected as the $1/z$ expansion treats the non-local hopping term perturbatively. In 2D, the comparison with the CM method works similarly well. 
The convergence of the method after a few iteration steps is demonstrated exemplarily in figure~\ref{convplots}.
The self-consistency scheme considerably improves the first and second order results of the $1/z$ expansion and converges rather fast. 
In the following two sections we apply this technique to study the gas-liquid transition in the BHM and the antibunching--bunching transition in the XYM.

\section{Bose-Hubbard model: Gas--liquid transition\label{sec4}}

\begin{figure}[b]
\centering
\includegraphics[width=0.9\textwidth]{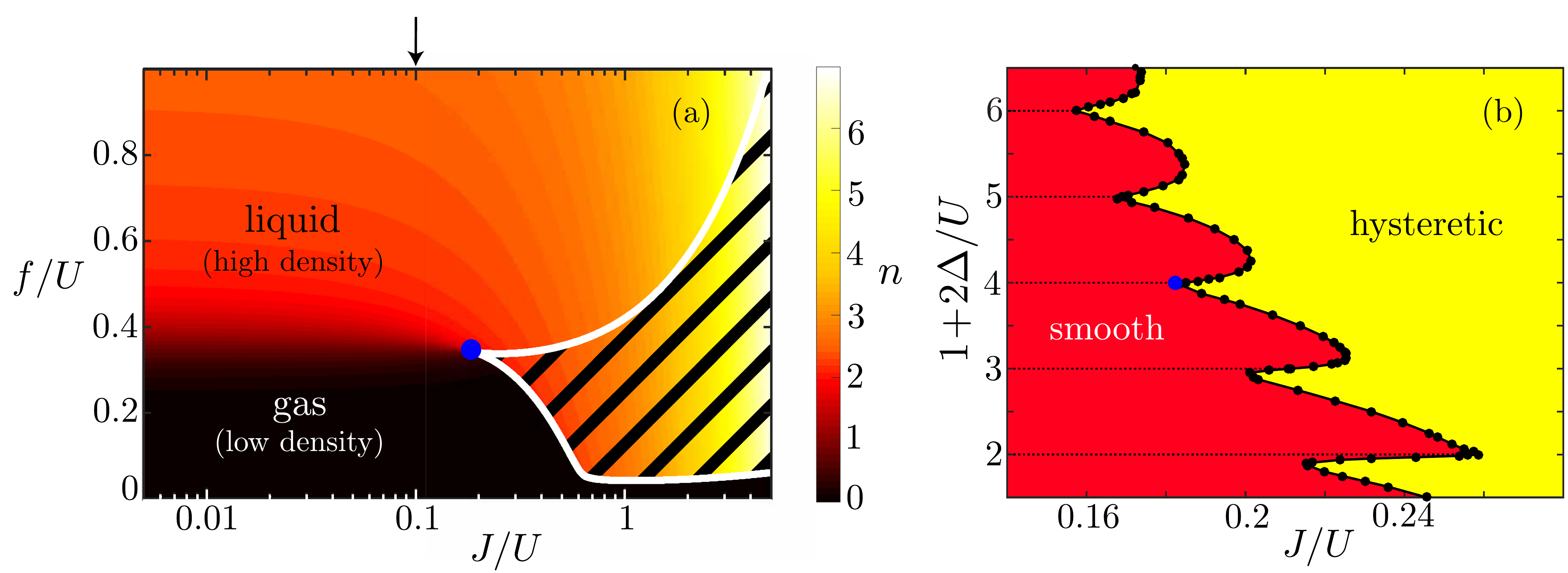}
\caption{(a) Mean-field density $n$ in color scale as a function of hopping $J/U$ and
drive strength $f/U$, illustrating the change
from the gas--liquid crossover at small hopping to the gas--liquid transition beyond the critical hopping $J_c$ (blue point). 
The region bounded by the white lines marks the bistable region of the mean-field theory (stripes
refers to densities of gas and liquid). The vertical arrow indicates the value $J/U=0.1$ chosen in figure~\ref{fig:Kvsf}.
(b) Boundary separating smooth from hysteretic gas--liquid transitions as driven by increasing 
the pump amplitude $f$, showing the quantum commensuration effect 
at successive $m$-photon resonances of the individual cavities, i.e., when $1+2\Delta/U=m$ assumes an integer value.
Figure adapted with permission from \cite{Biondi2016}.\label{fig:gas-liquid}}
\end{figure}

In this Section, we study the gas--liquid transition of photons as described by
the driven-dissipative Bose-Hubbard model \cite{boite2013,boite2014,Weimer2015}. 
The gas (liquid) phase is characterized by low (high) photon densities of the nonequilibrium steady-state.
The transition between the two phases can be driven by the coherent 
pump parameter $f/U$ at fixed detuning $\Delta/U$. 
For a single cavity, an exact solution provides a smooth crossover between the two phases when the pump strength is increased \cite{Drummond1980}.
In the lattice case, decoupling mean-field theory predicts that the gas--liquid crossover transforms into a 
hysteretic transition beyond a critical value of the intercavity hopping $J = J_c$.
The phase-diagram in the $f-J$ plane (at fixed detuning $\Delta/U$) is shown in figure~\ref{fig:gas-liquid}(a) with the critical point (blue) at $J_c$. 
Interestingly, one finds that the critical hopping $J_c$ is modulated as a function of the detuning $\Delta/U$
and exhibits a series of lobes, see figure~\ref{fig:gas-liquid}(b). The lobe structure is a manifestation of a quantum commensuration effect 
which favors the hysteretic transition over a smooth crossover whenever the drive frequency corresponds to a $m$-photon
resonance at $1 + 2\Delta/U = m$ \cite{Biondi2016}. 

Unfortunately, the lobe structure is particularly hard to calculate with exact numerical methods, because it requires
a high single-cavity photon number cutoff $n_p$ to capture the physics of multi-photon resonances.
This is why quantum trajectory simulations in Ref.~\cite{Biondi2016} were initially limited to 6 sites. 
However, despite the small system size, these simulations strongly substantiate the mean-field prediction: 
below the critical point ($J > J_c$), trajectories of each cavity switch independently and at random times between gas and liquid states; 
this behaviour changes drastically beyond the mean-field critical point ($J > J_c$), where all cavities of the array switch synchronously between gas and liquid phases.

\begin{figure}[t]
\centering
\includegraphics[width=0.8\textwidth]{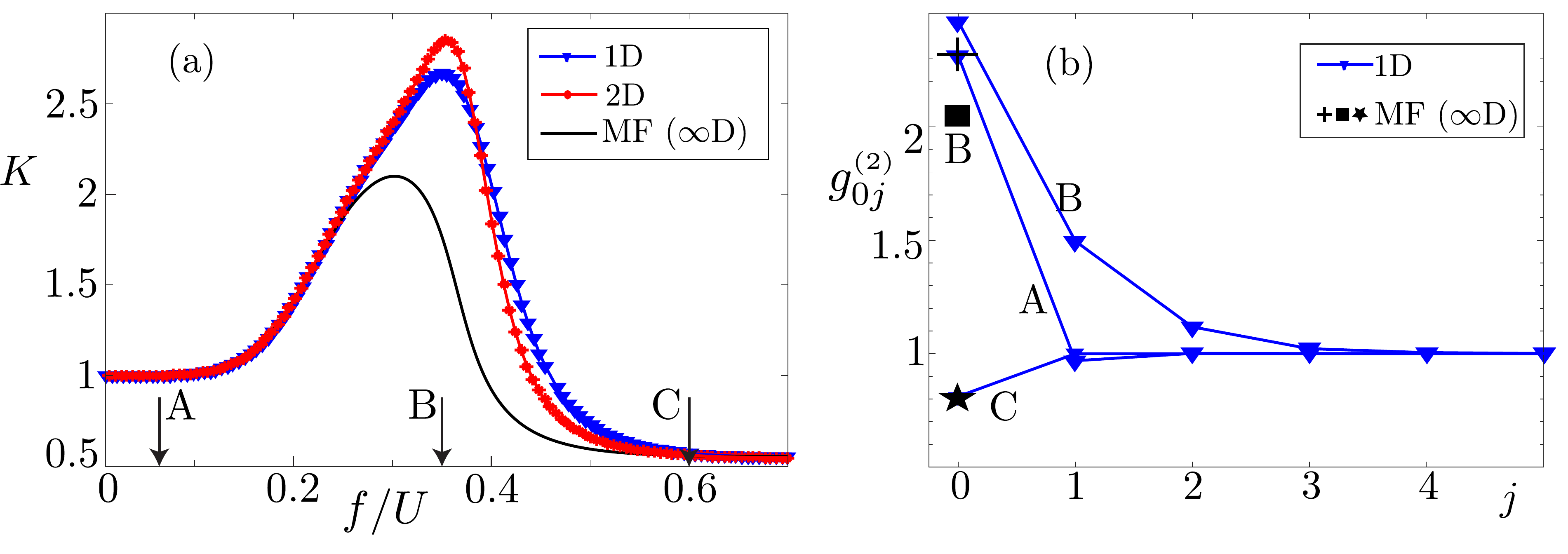}
\caption{Compressibility $K$ (a) versus 
drive $f/U$ at the 4-photon resonance $1+2\Delta/U=4$ for moderate hopping
$J/U=0.1$ and small dissipation $\kappa=U/20$. Shown are the mean-field ($\infty$D) results
as well as those of a first-order analysis in $1/z$ (see figure~\ref{fig_scheme}) 
including site-site correlations $\rho^c_{ij}$ for an array of 15 sites (1D) and a square lattice with $5\times5$ sites (2D). 
The compressibility exhibits a maximum in the gas--liquid crossover region, where it is largely enhanced
by site-site correlations. Panel (b) shows the pair-correlator
$g^{\scriptscriptstyle (2)}_{0j}$ evaluated in the gas (point A in (a)) and liquid (C) phases as well as in the crossover region (B). \label{fig:Kvsf}}
\end{figure}

In the following, we take a closer look at the gas--liquid transition and analyze compressibility and spatial correlations of the steady-state beyond mean-field using the $1/z$ expansion described in the previous section. 
First, we study density fluctuations via the compressibility
\begin{eqnarray}
\label{ktil_def}
   K = \frac{\langle\mathcal{N}^2\rangle - \langle\mathcal{N}\rangle^2}{\langle\mathcal{N}\rangle} = 1 + n\sum_{j=0}^{N-1} \left(g^{\scriptscriptstyle (2)}_{0j} - 1\right)\,.
\end{eqnarray}
Here, $N$ is the number of lattice sites, $\mathcal{N}=\sum_i a^\dagger_i a_i$ is the photon number operator and $g^{\scriptscriptstyle (2)}_{0j}$ is the second order 
coherence \eref{eq:g2}. Figure~\ref{fig:Kvsf}(a) shows the compressibility $K$ as a function of drive $f/U$ at fixed 
detuning $1+2\Delta/U=4$ and hopping $J/U=0.1$, i.e., corresponding to a vertical cut left from the critical point at $J_c \approx 0.18U$ in figure~\ref{fig:gas-liquid}(a). 
At weak drive $f/U\ll 1$ (gas phase), we find $K\approx 1$ as predicted by the mean-field approximation (solid line) and the $1/z$ expansion. 
Consequently, the gas phase is well described by a spatially uncorrelated, coherent state with $g^{\scriptscriptstyle (2)}_{0j}\approx 1$ for all sites $j$.
At large drive $f/U\gg 1$ (liquid phase), the prediction of the mean-field approximation, i.e., $K\approx 1/2$, also agrees well with the results obtained from the $1/z$ expansion.
In fact, the value $K\approx 1/2$ can be derived analytically from the single-cavity limit ($J=0$), where $g^{\scriptscriptstyle (2)}_{00}\approx1-1/m$ and $n\approx m/2$ at the $m$-photon resonance \cite{boite2014}.
We conclude that the effect of the lattice dimension is marginal deep in the gas and liquid phase, where the physics is well described by mean-field theory. 
However, the crossover region is characterized by strongly enhanced density fluctuations beyond mean-field. In particular, quantum fluctuations due to $1/z$ corrections in 1D as well as 2D 
strongly increase the compressibility with respect to the mean-field result. 
We attribute these enhanced fluctuations to the impending bistable behavior, see also \cite{Wilson2016}. 
Our $1/z$ results are also consistent with the quantum trajectory calculations in \cite{Biondi2016}, which show that synchronization effects already appear below the critical mean-field value $J_c$.

Making use of the $1/z$ expansion we also calculate the spatial correlation functions of the NESS for $J<J_c$. 
Figure~\ref{fig:Kvsf}(b) shows results for the pair-correlator
$g^{\scriptscriptstyle (2)}_{0j}$ in a 1D array for the
drive strengths $f/U$ indicated by the arrows in figure~\ref{fig:Kvsf}(a). At the 
compressibility peak ($f/U = 0.35$, line B), we find that bunched correlations
($g^{\scriptscriptstyle (2)}_{0j}>1$) extend further out in the lattice with a larger correlation length,
signaling the crossover between gas and liquid phases. 
This clustering of excitations is consistent with the coherent 
super-cavity formation as revealed by the quantum trajectory simulations in \cite{Biondi2016}. 
Away from the compressibility peak (A and C) photons at different sites are mostly
uncorrelated. The symbols at $j = 0$ indicate the mean-field values of the
on-site correlator $g^{\scriptscriptstyle (2)}_{00}$, which significantly
differ from the 1D results only at $f/U = 0.35$ (B, square). Similar outcomes
are obtained for the 2D lattice. We note, that the local Hilbert space cutoff $n_p$ (maximum photon number per cavity) 
required in figure~\ref{fig:Kvsf} is $n_p=6$, which would imply a huge
Liouvillian operator of size $M\approx 10^{21}$ in ED for the 2D case with $5\times5$ sites.

In summary, in this section we have shown with the $1/z$ method that bunched site-site correlations extend over many lattice sites
and largely enhance density fluctuations in the gas--liquid crossover regime of the driven-dissipative Bose-Hubbard model \eref{h_BHM}. The low
computational cost of the method allowed us to obtain insight for large lattices in 1D and 2D also in a regime of large photon numbers.
Unfortunately, it is difficult to analyze the hysteretic transition within the
$1/z$ expansion since the self-consistent approach does not always converge in this region of the phase diagram. 
In the next section, we will rather focus on the strongly-correlated regime $U\rightarrow \infty$ 
where the BHM \eref{h_BHM} is mapped to the spin-$1/2$ XYM \eref{h_XYM}.

\section{Spin-$1/2$ XY model: antibunching--bunching transition\label{sec5}}

In this section, we investigate the driven-dissipative spin-$1/2$ XYM in \eref{h_XYM}. 
In particular, we study the antibunching--bunching transition of the nearest neighbour correlator as a function
of the detuning $\Delta$, which was recently predicted in Ref.~\cite{MendozaArenas2015} using large scale MPS simulations.
In the following, we (i) provide a simple and analytic explanation of the transition based on a minimal model of two coupled spins (dimer),
(ii) reproduce exact numerical results with the self-consistent $1/z$ method to high accuracy and (iii) go beyond the MPS method by also studying the 2D case.

\begin{figure}[b]
\centering
\includegraphics[width=0.45\textwidth]{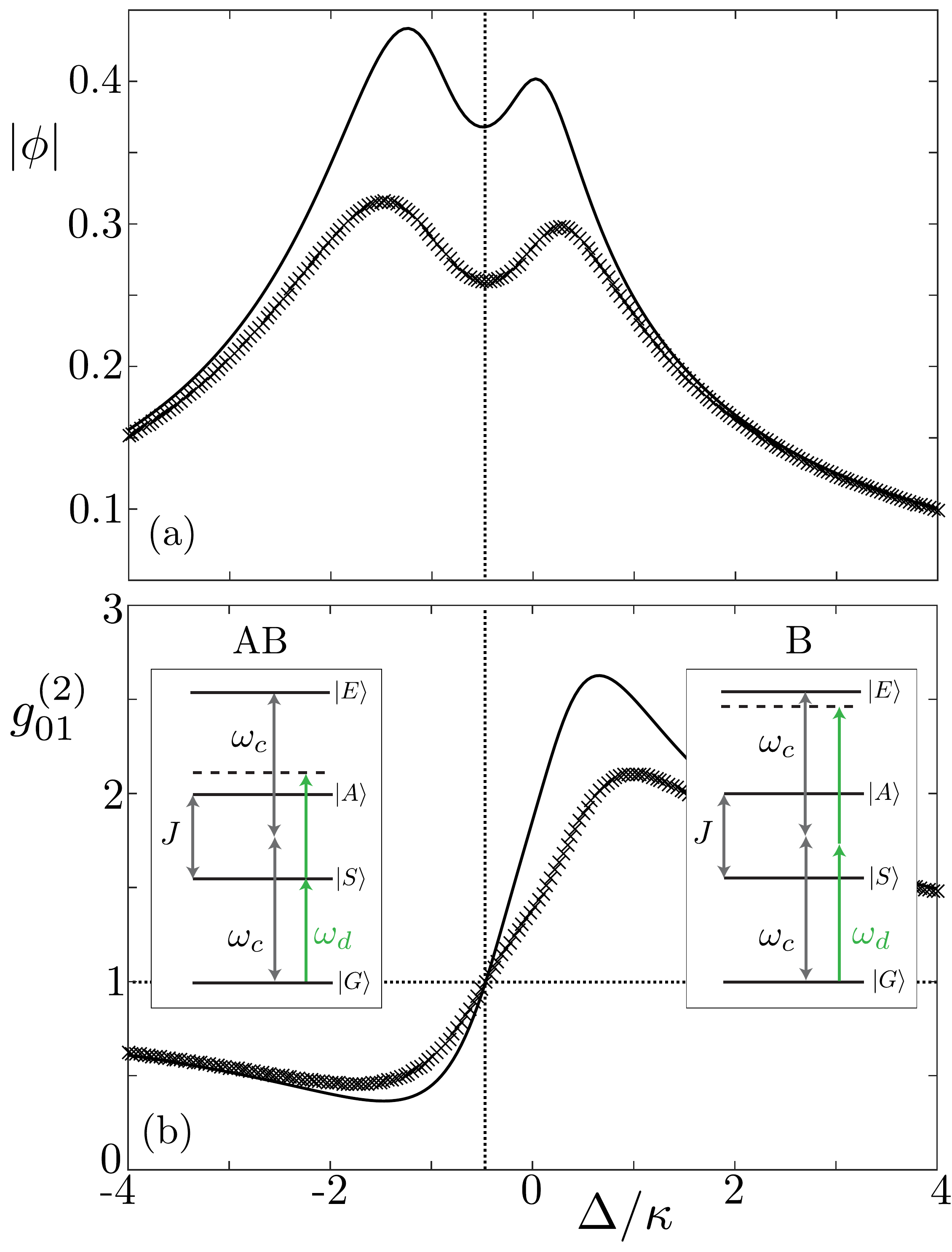}
\caption{Photon amplitude $|\phi|$ (a) and nearest-neighbor correlator
$g^{\scriptscriptstyle (2)}_{01}$ (b) for a dimer of two coupled spins as described by the XYM \eref{h_XYM}. 
Each spin is coherently driven with strength $f$ and decays with a rate $\kappa$. 
The panels display the dependence on drive frequency $\Delta/\kappa$ at fixed hopping $J/\kappa = 1.85$ 
and drive strength $f/\kappa = 0.5$, as obtained from ED (symbols) and an analytical $f/\kappa$ expansion (solid lines), see~\eref{eq:phiapp} and~\eref{eq:g2corr}.
In correspondence with the antiresonant behavior in $|\phi|$ (see also text) the nearest-neighbor correlator $g^{\scriptscriptstyle (2)}_{01}$ 
undergoes an antibunching--bunching crossover. The vertical dotted line marks the value $\Delta = -J/4$ where the correlator crosses unity. 
The insets named ``AB'' and ``B'' illustrate the level schemes of the dimer and the relevant transitions depending on the value of the detuning $\Delta = \omega_d - \omega_c$. When the drive is resonant 
with the symmetric superposition state of the dimer (see also text) the correlator is antibunched (``AB'') and the homodyne signal $|\phi|$ exhibit 
an antiresonant shape when the transition $\ket{G} \rightarrow \ket{S}$ starts to saturate; the correlator becomes bunched when the excited state of the dimer can be reached via two-photon transitions 
(``B''). The antisymmetric superposition state is dark and does not couple to the drive. \label{fig:b_ab_trans}}
\end{figure}

Before considering large lattices in 1D and 2D, it is instructive to focus on a simpler model consisting of a dimer of two coupled, driven-dissipative spins, i.e., 
a system described by the XYM in \eref{h_XYM} with $N=2$ sites and the associated four basis states $\{\ket{gg},\ket{ge},\ket{eg},\ket{ee}\}$.
Figure~\ref{fig:b_ab_trans} shows the photon amplitude (homodyne signal) $|\phi| = |\langle \sigma_0^- \rangle|$ 
and the nearest-neighbor correlator $g^{\scriptscriptstyle (2)}_{01} = \langle \sigma^+_0  \sigma^+_{1}
\sigma^-_0 \sigma^-_{1}\rangle/\langle \sigma_0^+ \sigma_0^-
\rangle^2$ as a function of detuning $\Delta/\kappa$ at fixed drive strength $f/\kappa=0.5$ and hopping $J/\kappa = 1.85$. 
The exact diagonalization results (symbols) are qualitatively reproduced by an approximate solution
of the master equation \eref{lindblad_full}, which we have obtained by expanding the density matrix elements perturbatively in powers of $f/\kappa$.
For the homodyne signal we obtain for weak drive powers
\begin{eqnarray}
|\phi| \approx  \frac{f}{\sqrt{A}(1 + 2f^2/A)}\left(1 + \frac{f^2}{\Delta^2 + \kappa^2/4}\left(1 - \frac{J(J/2 + \Delta)}{2A}\right) \right)
\label{eq:phiapp}
\end{eqnarray}
and for the second-order correlation function
\begin{eqnarray}
g^{\scriptscriptstyle (2)}_{01} \approx 1 + \frac{J}{2}\frac{J/2 + 2\Delta}{\Delta^2 + \kappa^2/4}\left(1 - \frac{2f^2}{\Delta^2 + \kappa^2/4}\right)
\label{eq:g2corr}
\end{eqnarray}
with $A = (\Delta + J/2)^2 + \kappa^2/4$. The analytic results \eref{eq:phiapp} and \eref{eq:g2corr} correspond to the solid lines in figure~\ref{fig:b_ab_trans}.  
We note that a quantitative agreement between analytic and exact results is achieved for smaller pump strength  $f/\kappa \lesssim 0.2$. 
Simple algebra reveals that \eref{eq:g2corr} changes from antibunched to bunched when $\Delta \approx - J/4$. 
Interestingly, the splitting of the resonance peak in the homodyne signal occurs at a similar value.
The resulting antiresonant lineshape of the homodyne signal is a signature of photon blockade \cite{imamoglu1997}.
It is well known from the Jaynes-Cummings model \cite{bishop2009,nissen2012}, 
where it is usually referred to as the `dressing of the dressed states' \cite{Shamailov2010} and can be explained
by the optical Bloch equations \cite{Arecchi1965}. Such a nonlinear effect 
arises under strong pumping due to the saturation of the transition between the ground and
an excited state of the system. The antiresonance is peculiar to the homodyne/heterodyne detection scheme measuring the photon
amplitude $|\phi|$ rather than the photon density $n$. The latter only exhibits power broadening when the drive strength increases. 
Recently, the properties of the antiresonance in coupled qubit-cavity arrays was studied in \cite{nissen2012}. 

We now argue that the antibunching--bunching crossover as well as the antiresonance
can both be understood in terms of the relevant eigenstates of the dimer model (inset in figure~\ref{fig:b_ab_trans}(b)):
When the drive is resonant with the symmetric superposition $\ket{S} = (\ket{ge} + \ket{eg})/\sqrt{2}$, a saturation of the transition $\ket{G} \rightarrow \ket{S}$  
leads to the antiresonant shape in $|\phi|$ and more antibunched correlations. A simultaneous excitation of both spins is not possible (see level scheme in the inset named ``AB'').
Increasing the drive frequency beyond $\Delta = - J/4$ allows to populate more efficiently the excited state $\ket{E}=\ket{ee}$ via a two-photon transition. This leads to a bunching of excitations
in neighboured cavities (inset named ``B''). Note that the antisymmetric state $\ket{A} = (\ket{ge} - \ket{eg})/\sqrt{2}$ is dark and does not couple to the drive.

\begin{figure}[b]
\centering
\includegraphics[width=1\textwidth]{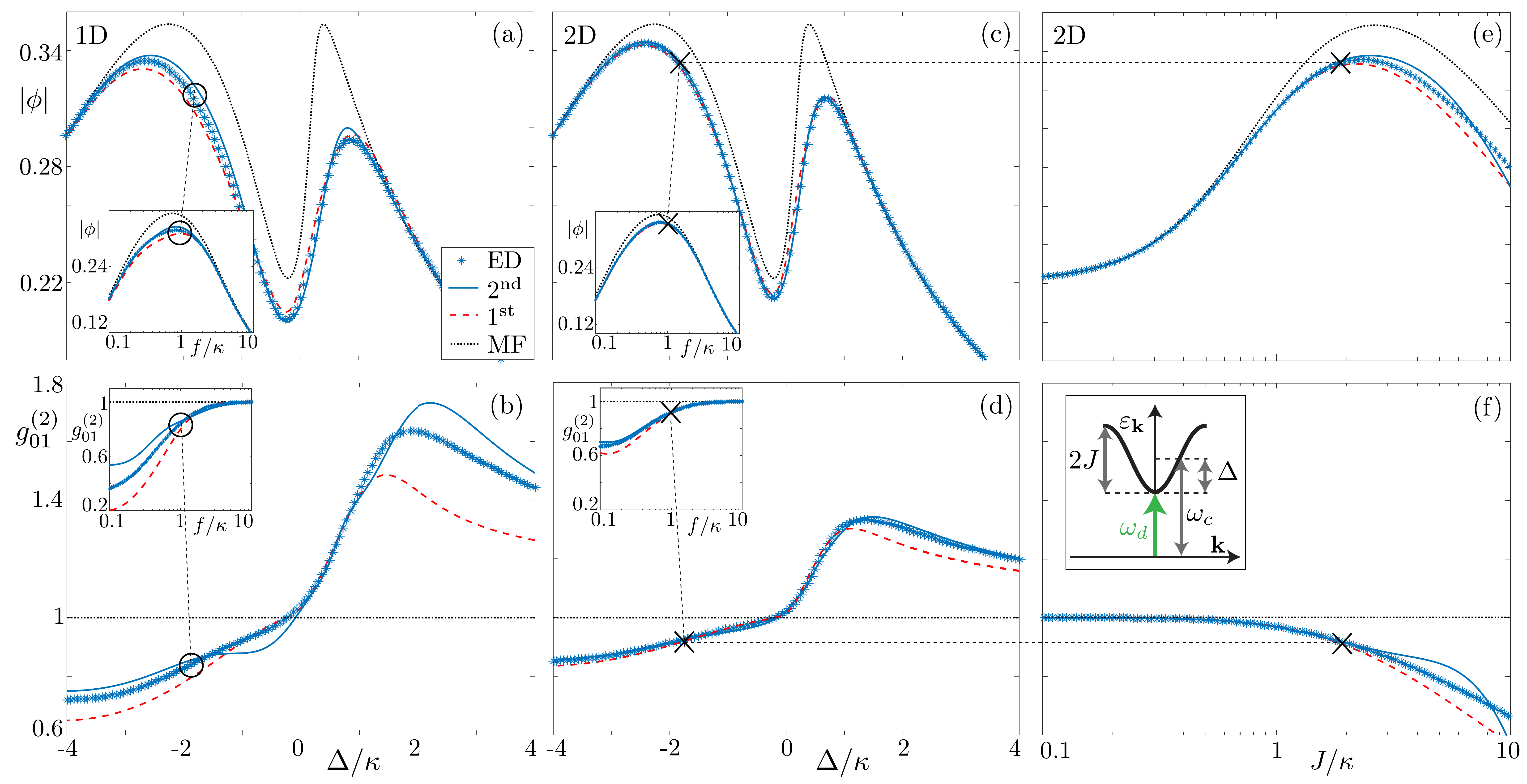}
\caption{Photon amplitude $|\phi|$ [(a),(c),(e)] and nearest-neighbor correlator
$g^{\scriptscriptstyle (2)}_{01}$ [(b),(d),(f)] for the XYM \eref{h_XYM} on a
1D array of 9 sites [(a) and (b)] and a 2D (square) lattice of $3\times3$
sites [(c) and (f)]. Panels (a)--(d) display the dependence on drive frequency
$\Delta/\kappa$ at fixed hopping $J/\kappa = 1.85$ and drive strength
$f/\kappa = 1$. The insets in (a)--(d) show the dependence on drive strength
at fixed frequency $\Delta = - J$. The circles (crosses) in 1D (2D) mark the
corresponding points in different plots. Panels (e) and (f) show the
observables as a function of hopping $J/\kappa$ at fixed drive strength
$f/\kappa = 1$ when the drive frequency is kept resonant with the bottom of
the photon band ($\Delta = - J$) as chosen in the inset. In all the panels,
the curves correspond to different orders in the $1/z$ expansion while symbols
are results from exact diagonalization (ED). \label{testED}}
\end{figure}

We now increase the system size and discuss both phenomena in large lattice systems using the $1/z$ expansion
and ED. In figure~\ref{testED}, we choose a moderate hopping $J/\kappa=1.85$ and
vary the drive frequency $\Delta/\kappa$. As with the dimer, we observe a pronounced antiresonance 
in the photon amplitude $|\phi|$ together with a changeover from antibunching to bunching in the correlator
$g^{\scriptscriptstyle (2)}_{01}$. As already observed in table \ref{tables}, the
largest deviations occur between the Gutzwiller mean-field-
(MF, black dotted line) and the first-order results of the expansion ($1^{\rm
st}$, red dashed line). The latter reproduces the exact
numerical data well. This can be attributed to the local nature of drive and dissipation, 
which limits correlation effects to a few lattice sites \cite{jin2013,biondi2015}. 
In 2D, the method captures the local
photon amplitude $|\phi|$ as well as the nonlocal correlator $g^{\scriptscriptstyle (2)}_{01}$
more accurately than in 1D. The corrections to the MF results
become smaller with increasing lattice dimension.
We also performed simulations in 3D (not shown), which confirm these general statements.

The insets in figure~\ref{testED}(a)--(d) display the dependence on the drive
strength $f$ at fixed drive frequency $\Delta = - J$. 
Again, we find excellent agreement between exact results and the $1/z$ expansion in 2D. 
Note that in 1D the expansion is slightly less
accurate in describing site-site correlations.
Panels (e)--(f) of figure~\ref{testED} display the hopping dependence of the
observables in 2D when the drive is kept resonant with the bottom of the
photon band ($\Delta = - J$) as illustrated in the inset. Already at small
hopping $J/\kappa \approx 0.6$ the MF value of $|\phi|$ shown in (e) deviates
from the exact result, approximately when the nearest-neighbor correlator
shown in (f) departs from unity. The $1/z$ expansion performs better in
reproducing the correct local as well as the nonlocal observables up to
$J/\kappa \approx 5$, i.e., for roughly one order of magnitude larger values
of $J/\kappa$.

In summary, we have demonstrated that the $1/z$ expansion can reproduce the antibunching--bunching transition in 1D
previously analyzed with more demanding computational techniques such as MPS \cite{MendozaArenas2015}. Furthermore, 
larger lattice dimensions (2D and 3D) --- currently out of reach for MPS-based approaches --- can easily 
and accurately be studied with our method. We have also provided a simple analytic argument based on a dimer model of two coupled spins,
which explains the physical origin of the antibunching--bunching transition.

\section{Summary\label{sec7}}
In summary, in this paper we have developed a self-consistent scheme based on a $1/z$ expansion 
with the goal of studying efficiently the nonequilibrium steady-state of correlated photons in cavity arrays beyond the mean-field 
approximation.
Going to second order in the $1/z$ expansion in 1D and 2D, we have included up to
three-site correlations in our analysis and have obtained accurate agreement
with exact numerical methods, particularly in
the small to moderate hopping regimes. We have studied two applications in the context of the driven-dissipative BHM and XYM,
which testify that this $1/z$ expansion represents a valuable tool that confirms the qualitative correctness of the mean-field results and provides quantitative improvements
on the theoretical predictions. The approach can be easily applied to a large variety of nonequilibrium lattice systems and comes with a remarkably low computational cost, which makes it an appealing 
alternative to the few available methods for the simulation of interacting open systems in large lattice dimensions. \\

We acknowledge support from the National Centre of Competence in Research `QSIT--Quantum
Science and Technology' (MB) and the US Department of Energy, Office of Basic
Energy Sciences, Division of Material Sciences and Engineering under Award
No.\ DE-SC0016011 (HET).

\end{document}